\documentstyle[11pt,pazh2,psfig]{article}

\newcommand{\km}{\,\mbox{km}\,\mbox{s}^{-1}}
\def\Ha{\hbox{H$_\alpha$~}}

\begin{document}

\title{The Effect of Projection on the Observed Gas Velocity Fields
in Barred    Galaxies\thanks{Astronomy    Letters,    Vol.26,No~9, 2000,
pp.565-571.   Translated  from  Pis'ma  v  Astronomicheskii
Zhurnal, Vol.26, No. 9, pp. 657-664. Translated by A.Dambis.}}

\author{Moiseev A.V. \inst{a}
\thanks{E-mail address for contacts: moisav@sao.ru}
\and Mustsevoi V.V. \inst{b}}

\institute{
\saoname
\and Volgograd State University, Volgograd, Russia
}
\maketitle

\begin{abstract}
\small
The problem of determining the pattern of gas motions in the central
regions of disk spiral galaxies
is considered. Two fundamentally different cases -- noncircular motions in
the triaxial bar potential and motions
in circular orbits but with orientation parameters different from those of
the main disk -- are shown to have similar
observational manifestations in the line-of-sight velocity field of the
gas. A reliable criterion is needed for
the observational data to be properly interpreted. To find such a
criterion, we analyze two-dimensional nonlinear
hydrodynamics models of gas motions in barred disk galaxies. The gas
line-of-sight velocity and surface
brightness distributions in the plane of the sky are constructed for
various inclinations of the galactic plane to
the line-of-sight and bar orientation angles. We show that using models of
circular motions for inclinations
$i > 60^\circ$ to analyze the velocity field can lead to the erroneous
conclusions of a ``tilted (polar) disk'' at the circumnuclear region.
 However, it is possible to distinguish bars from tilted disks by
comparing the mutual orientations of the
photometric and dynamical axes. As an example, we consider the velocity
field of the ionized gas in the galaxy NGC 972.
\end{abstract}

\section{Introduction}
The analyses of the structure and kinematics of spiral
galaxies often require detailed information on the
gas motion patterns in the stellar systems considered:
the rotation curve and whether the velocity field
includes noncircular motions. If gas motions are circular
it is enough to measure line-of-sight velocity profiles
along two sections passing through the disk center
to unambiguously infer the orientation parameters and
determine the rotation curve (Zasov 1993). This however
only a  rough approximation of the real
dynamics of the gaseous disk: the velocity fields of spiral
galaxies can exhibit large-scale noncircular motions
amounting to $20-30 \km$ , which are due by the spiral pattern.

Much more detailed information can be obtained by
measuring the full velocity field, i.e., by determining
the line-of-sight velocity distribution over the entire
galaxy disk. This has since long become common practice
among radio astronomers who use interferometry
technique to measure the spectral lines of neutral or
molecular gas. Similar results are obtained at optical
wavelengths with 2D-spectroscopy using Fabry-Perot
interferometers or integral field spectrographs. The
common feature of all these methods is that they all
involve construction of the so-called ``data cube'' (Tully
1974). A special technique allows the observational
data to be reduced to a form where each image element
has its own individual emission- (absorption-) line
spectrum. The line-of-sight velocity field is then constructed
from the Doppler shifts of the lines studied.

It is, however, often difficult to unambiguously
reconstruct the rotation curve and analyze the direction
of noncircular motions in the galaxy even if the full
line-of-sight velocity field is available. Noncircular
motions, which are primarily due to the spiral pattern,
make it impossible to isolate the circular rotation component
simply by averaging the total gas velocity field.
Fourier analysis of the line-of-sight velocity field (Lyakhovich
et al. 1997) allows the full spatial pattern
of gas motions in a spiral galaxy to be reconstructed
under certain assumptions about the nature of the spiral
structure.

Barred galaxies exhibit noncircular motions of even
larger amplitude. Although the interpretation of observed
gas velocities in galaxy bars remains highly ambiguous,
a number of important conclusions to be made
about the motion pattern in the bar can be drawn by
analyzing the gas rotation in terms of the circular-motion
model. The bar produces a turn of the dynamical
axis (i.e., the line of maximum line-of-sight velocity
gradient) relative to the line-of-nodes (Chevalier and
Furenlid 1978; Afanasiev et al. 1992; Zasov \& Moiseev
1999) and this turn can be determined using the
well-known algorithms for the analysis of disk galaxy
velocity fields.

The most comprehensive conclusions about
a bar in a  galaxy  can be drawn
from numerous works on nonlinear computer simulations
involving hydrodynamics equations written for
``gas'' consisting of macromolecules in the form of gas
clouds. See Lindblad (1996) for a review of various
model simulations. The common result of these works
is that a bar to be an efficient mechanism
for removing the angular momentum of the gas,
thereby making the latter to partly move centerward
along the bar and lose the energy at its shock edges
(Athanassoula 1992a, 1992b; Levy et al. 1996). This
motion pattern distorts the observed velocity contours,
aligning them along the bar, whereas the dynamic axis,
which is perpendicular to the isovelocities  it
crosses, must, accordingly, turn in the opposite

We must, nevertheless, bear in mind that the turn of
the dynamic axis can also be caused by a tilted disk at
galaxy center (Zasov \&  Sil'chenko 1996; Zasov
\&  Moiseev 1998). However, in this case, the
 axis of a circular rotating disk coincides with the
line-of-nodes. Also the photometric and dynamic
axes should therefore turn in the same direction. These
are only qualitative considerations; therefore, concrete
computations are needed, because the projection effect
with radial gas flows in the bar can
produce a pattern similar to that expected in the
tilted disk case.

Extensive panoramic spectroscopy of HII velocity
fields in spiral galaxies have been obtained with the
Fabry-Perot interferometer attached to the 6-m BTA
telescope of the Special Astrophysical Observatory of
the Russian Academy of Sciences. It is very important,
when interpreting these data, to analyze gas motions in
the bars and identify eventual minibars (smaller than
1 kpc) from the dynamical manifestations they cause in
the velocity field (see, e.g., Afanasiev et al. 1989;
Zasov \& Sil'chenko 1996; Sil'chenko et al. 1997). To
investigate whether the above patterns in the behavior
of the mutual orientations of photometric and dynamical
axes can in principle be applied to analyze
observational data, it would be useful to consider the results of
numerical velocity-field simulations with allowance for
sky-plane projection effects and limited spatial and
spectral resolution.

In this paper, we use the initial data, which are similar
to the results of computations by Levy et al. (1996),
to construct simulated HII velocity fields similar
to those actually observed in the \Ha and [NII] lines with the
standard Fabry-Perot attached to the BTA telescope
(see Dodonov et al. (1995) for a detailed description of
the instrument). These data have a spatial resolution of
$(1.5-4)''$  and a spectral resolution $(50-150) \km $
in velocity terms, with an accuracy of inferred line-of-sight
velocities of $(3-10) \km$ .

\section{Velocity fields constructing}

Our simulations are based on the numerical solution
of hydrodynamics equations in an external gravitational
field. The adopted gravitational potential model
consists of a nonaxisymmetric perturbation superimposed
on an axisymmetrical component (bulge, disk, and
halo) see Matsuda et al.(1987).

The nonaxisymmetric perturbation (stellar bar
mode) is turned on smoothly and gradually increased
until it reaches the fixed level. The model exhibits a
specific quasi-periodic regime of gas passage through
the bar (Levy et al. 1994, 1996). Interestingly, the
resulting flow pattern is bisymmetric—the gas flows
centerward along the bar in two sectors, whereas in the
other two sectors the flow direction is reversed; i.e., the
gas flows away from the disk center due to the saddle
point of the gravitational potential. The developing
ar also produces a two-armed spiral pattern
outside the corotation circle of the bar.

We used the results of simulations identical to those
described in detail by Levy et al. (1996). The only difference
was that we party adjusted the gravitational
potential to closely reproduce the rotation curve of our
Galaxy (Haud, 1979). Our model had a maximum linear
rotation velocity of $250 \km$ ; disk scale length of
3 kpc; radial bar scale length of 1 kpc, and a radius of
a simulated region 10 kpc. We further assumed that the
galaxy is at a distance of 20.6 Mpc, where 1\arcsec~ corresponds  to 0.1
kpc, and has a systemic line-of-sight
velocity (i.e., the center-of-mass velocity relative to the
observer) of $V_{sys}= 1545\km$ .

Our numerical simulations yielded the distributions
of gas surface density $\sigma_{gas}(R,\varphi)$, radial $V_R(R,\varphi)$ and
azimuthal $V_\varphi(R,\varphi)$ gas velocity components. Here, $R$
and $\varphi$ are the radial and azimuthal coordinates in the
galactic plane, respectively. The polar grid cells had sizes
of $\Delta\varphi = 2^\circ$ in the azimuthal and $0.016<\Delta R<0.2$ kpc in
the radial direction, depending on the galactocentric
distance. Based on these data, we constructed model
cubes using the following algorithm.
Line-of-sight velocity at point $(R,\varphi)$:

\begin{equation}
 \begin{array}{l}
 V_{obs}(R,\varphi)=V_{sys}+V_{R}(R,\varphi)\sin\varphi\sin i\\
     \qquad {} +V_{\varphi}(R,\varphi)\cos\varphi\sin i
 \end{array}
\end{equation}

where $V_{sys}$ and $i$ are the systemic line-of-sight velocity
and the inclination of the galaxy plane to the sky plane,
respectively. $\varphi = 0$ set for the line-of-nodes.

We projected distributions $V_{obs}(R,\varphi)$ and
$\sigma_{gas}(R,\varphi)$ onto a Cartesian grid with a cell size of
0.02 kpc (0\farcs8), assuming that the position angle of the line of
nodes in the sky plane is $PA_0=90^\circ$. As a result, we
obtained the distributions of line-of-sight velocity and
gas surface density distributions $V_{obs}(x, y)$ and $\sigma_{gas}(x, y)$,
respectively. The emission-line spectrum at
each point is fitted by a Gaussian centered on $V_{obs}(x, y)$ with a
$FWHM = 130\km$, which approximately
corresponds to the spectral resolution of the
Fabry-Perot Interferometer attached to the BTA telescope.
We neglected the velocity dispersion of individual gas
clouds during a  construction of  the total spectrum because of the coarse
spectral resolution. We
assumed that the spectral-line intensity is proportional
to $\sigma_{gas} (x, y)$.  The images in all spectral channels
were Gaussian smoothed to make the resulting spatial
resolution equal to 2\arcsec, which is close to the typical
seeing value during real observations.

Based on the smoothed-cube spectra, we constructed
the velocity field and the spectral-line image
(the observed distribution of gas surface density) and
determined the observed orientation parameters of the
gaseous disk using the technique similar to that
described by Begeman (1989). A similar algorithm is
used in the well-known GIPSI radio astronomical data
reduction software to analyze the observed velocity
fields.

Below, we briefly describe the procedure used for
the analysis.

Introduce polar coordinate system $(r,PA)$ in the sky
plane, where r and PA are the distance from the rotation
center and position angle, respectively. The observed
line-of-sight velocity $V_{obs}$ in the case of purely circular
rotation is
\begin{equation}
\label{model}
 \begin{array}{l}
   V_{obs}(r,PA)=V_{sys}\\
  \qquad {}+V_{rot}(R(r))\frac{\cos(PA-PA_0)\sin i}{(1+\sin^2(PA-PA_0)\tan^2i)^{1/2}}
 \end{array}
\end{equation}

and the distance from the rotation center in the galaxy
plane is

 \begin{equation}
 \label{R}
 R(r)=r(1+\sin^2(PA-PA_0)\tan^2i)^{1/2}
 \end{equation}

where $V_{rot}$ and $PA_0$ are the circular rotation velocity and
position angle of the line-of-nodes, respectively.

During reduction, the observed velocity field is divided
into elliptical rings defined by equation (3) for
$R = const$. A nonlinear least squares technique are used
to fit model curve (2) to the observed dependence
$V_{obs} (PA)$. As a result, we obtain for each radius $r$ the
corresponding disk orientation parameters $PA$ and $i$,
and the circular rotation velocity $V_{rot}$.

Note that while the positions of the extremes of function
(2) allow PA to be determined quite unambiguously;
$i$ estimates are much more uncertain. Thus, at
small inclination angles $i$ (galaxies seen as face-on), it
is impossible to separate the contributions of $V_{rot}$ and $i$
to the observed velocity projection and only $V_{ROT}\sin i$
can be unambiguously inferred. Begeman (1989) used
a reduction of radio data to show the coupling of the
two parameters at inclination angles $i < 40^\circ$.

We constructed and analyzed simulated fields for
various bar development stages and inclination angles
ranging from 30\degr~ to 70\degr. We varied the angle between
the line-of-nodes and the bar with a step of 15\degr~ and constructed
a total of more than 400 simulated galaxies.

\section{Analysis of the results}

In Fig. 1, we plot the gas surface-density distribution
and line-of-sight velocity contours for various bar
orientation angles. This is a typical pattern
for dimensionless time instants $T < 3.4$ (where $T = 1$
corresponds to one bar revolution). The bar then saturates
and the spirals degenerate into a pseudo-ring (see
Levy et al. (1996) for details). The contour turns inside
$r < (10-15)''$ are caused by the bar, but on
greater galactocentric distances they are determined by
the motions in spiral arms. The typical S-shaped radial-velocity
contours can be seen in the central region. It is
interesting that the contours in question are distorted
even if the bar is aligned along the major axis of the galaxy
($\phi = 0$, where $\phi$ is the angle between the major axis
of the bar and the line-of-nodes measured in the galaxy
plane). The line-of-sight velocity projection is formally
equal to zero in this case; however, the bar has a finite
width and velocity perturbations contain an azimuthal
component, which also contributes to the radial velocity.

\begin{figure*}
\psfig{figure=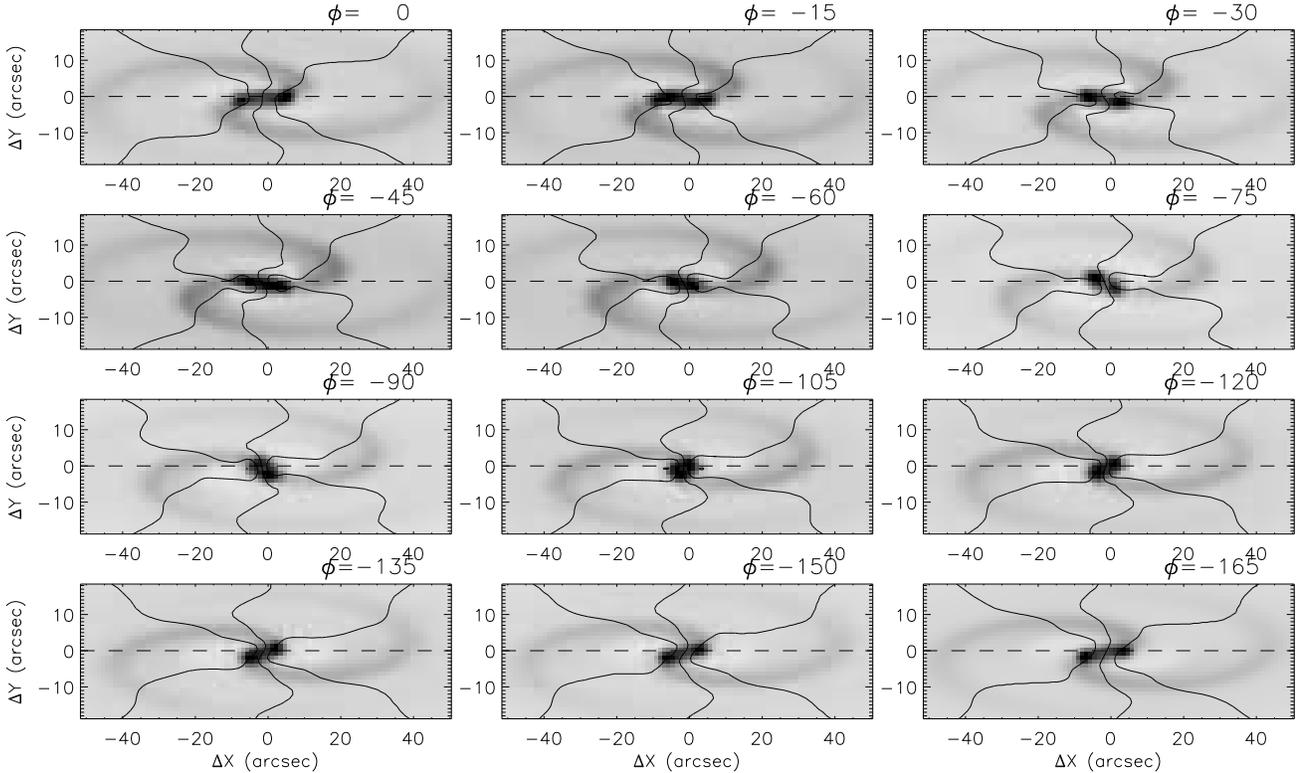,width=\textwidth}%<---Nii
\caption{Sky plane projection of the gas surface brightness in the central
region for various bar turn angles $\phi$ (indicated at the top right
corner in each graph) at time $T = 2.0$ since the start of the simulation.
Contours show the $V_{sys}$ and $V_{sys}\pm100\km$ line-of-sight
isolines of model velocity field. The galaxy plane tilt is $i = 70^\circ$; the
dashed curve shows the line-of-nodes (position angle $PA = 90^\circ$).
}
\end{figure*}

Fig. 2 shows typical results for dynamical axis
orientation for an inclination angle of $i = 70^\circ$.
 At the very center of the galaxy, the dynamical
axis always deviates significantly from the line of
nodes $PA_0 = 90^\circ$ (except $\phi\approx0^\circ$, $-90^\circ$ i.e., in the
vicinity of the major and minor axes, respectively).
Beginning with $r < 5''$, the angle between the two
lines decreases and becomes negligible at $r \approx 10''$
(in the vicinity of the bar tips). At greater galactocentric
distances, the dynamical axis turns to the other side of
the line-of-nodes and misalignment disappears only at
$r > 20''$.

As the bar evolves, a shock develops at its leading
edge and this shock is especially conspicuous in the
pressure maps due to the strong gas density in the region considered.
Such an arrangement of shock fronts is consistent
with both the observations of dust lanes in bars and
with the results of earlier numerical simulations (Athanassoula
1992a, 1992b). The beat-frequency modulation
about $PA_0$ is due to the complex velocity pattern at
the bar tips, where the shock front moves from the leading
bar edge to the inner edge of the trailing arms
(because the corotation radius is $r\approx 20''$). The $PA$
variation pattern is asymmetric with respect to $\phi=-90^\circ$,
because of a preferred direction of angle-counting
determined by the pitch angle of the spiral density
wave, whose direction is opposite to that of galactic
rotation (trailing spiral).

The thick line in Fig. 2 shows the orientation of the
sky-plane projection of the (the observed photometric)
axis of the bar. It is evident from the figure that, in spite
of the complex pattern of PA radial variation, the
dynamical axis in the bar $(r < 10'')$ always turns
with respect to the photometric axis in the direction
opposite that of the line-of-nodes: the two lines turn in
antiphase. This conclusion applies to all simulated
fields with different  $i$, $\phi$ and $T$.

In order to better illustrate this effect,
Fig. 3 shown the deviations of the dynamical axis (averaged
over all points with $r<6''$ depends on the bar
orientation. In this figure, the amplitudes of the turn of
the dynamical axis are somewhat smoothed by averaging;
however, the overall trend remains conspicuous at
all inclination angles and various bar development
stages.

\begin{figure*}
\psfig{figure=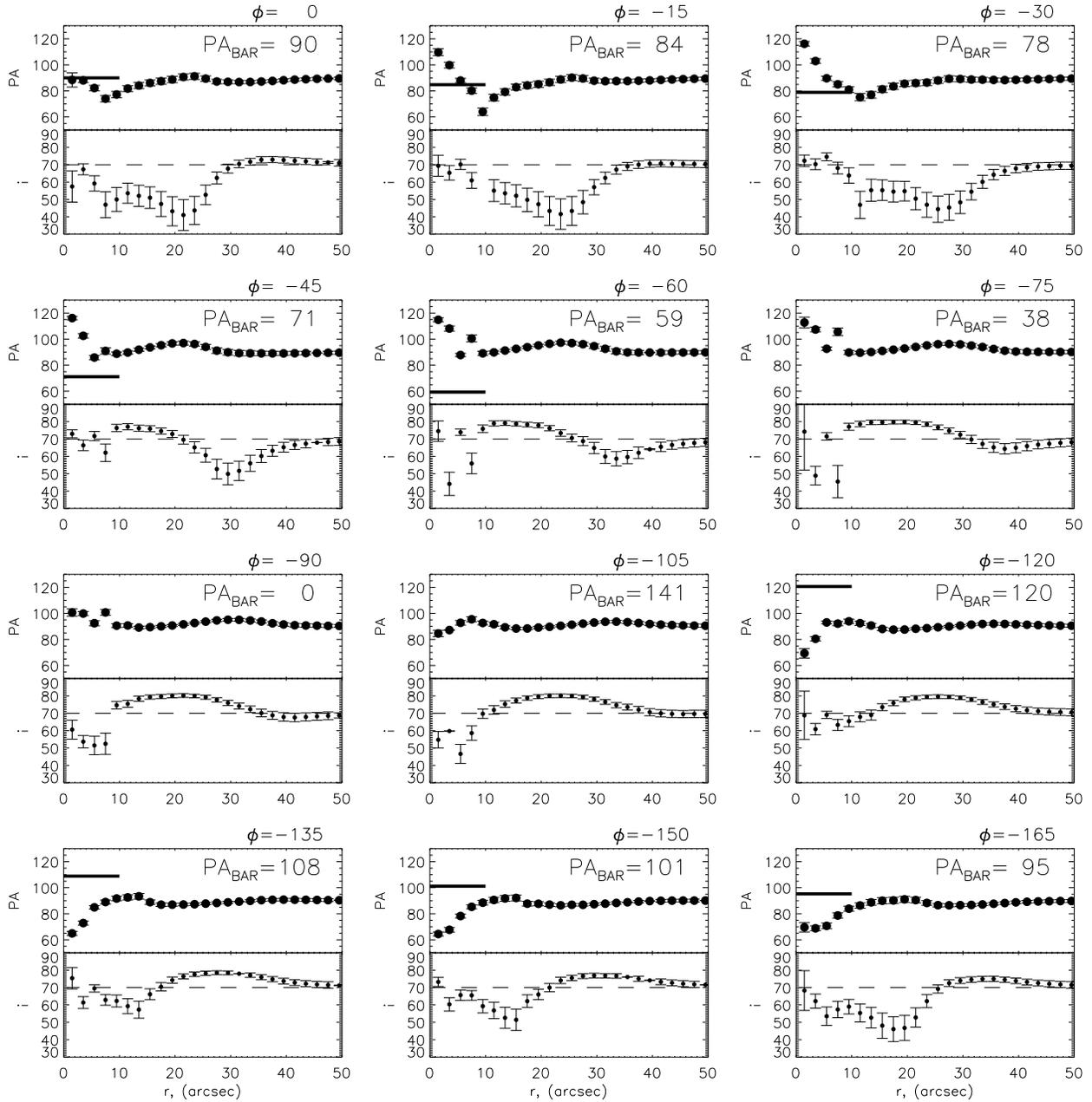,width=\textwidth}%<---Nii
\caption{
Results for model velocity fields corresponding to various bar
turn angles at time $T = 2$. Radial variations of the dynamical
axis orientation PA and tilt angle i. Solid lines in PA plots show the
position angle of the bar projected on the sky plane (indicated
by ``$PA_{bar}=...$'' on each graph). Dashed lines on $i$ plots show the
galaxy inclination $i = 70^\circ$
}
\end{figure*}

At $\phi$ angles ranging from $0^\circ$ to $\pm45^\circ$ (which correspond
to $PA\approx\pm(30^\circ-45^\circ)$ depending on the inclination
angle $i$), the dynamical axis turns virtually in antiphase
with the bar orientation. With a further increase of the
angle between the bar and the line-of-nodes the dynamical
axis begins to ``lag behind''; its orientation changes
slower than that of the photometric axis and it virtually
coincides with the line-of-nodes at $\phi\approx\pm 90^\circ$. However,
the turn of the dynamical axis from the line-of-nodes is
always in the sense opposite that of the bar axis. The
lag of the dynamical axis is a result of a combined
effect of projection and that the bar fails
to stop the gas rotation completely (the maximum bar
contrast in our experiment was 30\%), and there is
always an azimuthal component besides the radial
motions.

\begin{figure*}
\psfig{figure=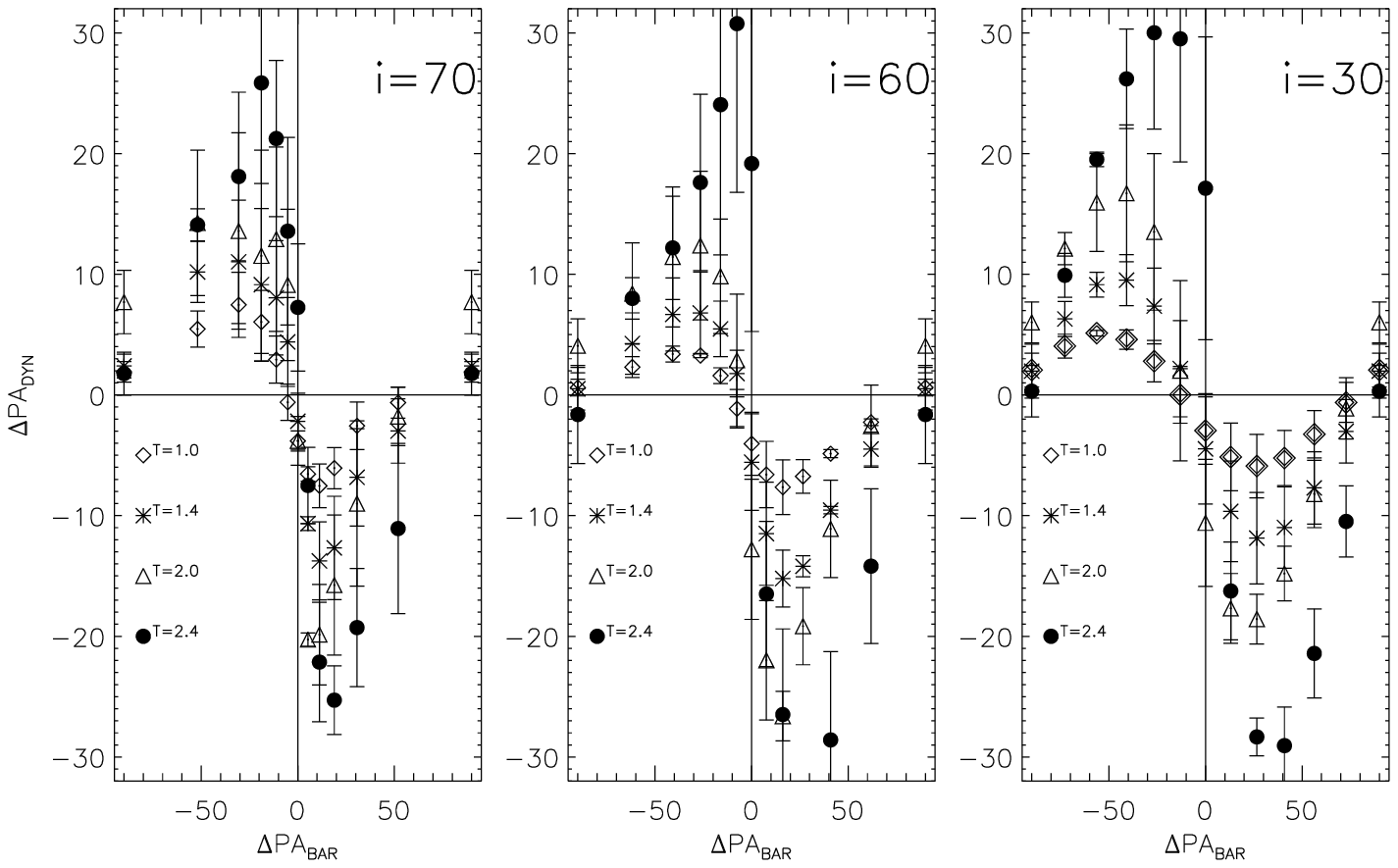,width=15 cm }%<---Nii
\caption{
Mean deviation of the position angle of the line-of-nodes in the bar region
$\Delta PA_{dyn}$ plotted as a function of the bar position
angle $\Delta PA_{bar}$ for  $i=70^\circ,60^\circ,30^\circ$. Different signs correspond to different
dimensionless time instants $T$.
}
\end{figure*}

We now analyze the behavior of the inclination $i$ formally
inferred from the velocity fields (Fig. 2). Radial
motions in the bar region distort relation (2), which
should have been observed if the rotation is purely circular.
Therefore the inclination inferred at these radii
can differ substantially from the true inclination of the
disk. The variations of $i$ can have amplitudes as high as
$20^\circ - 30^\circ$ and are often rather chaotic. However, the following
pattern can be observed: for galaxy inclinations $i \ge 60^\circ - 70^\circ$,
the formally inferred inclination in the
bar region becomes virtually constant at $\phi > -30^\circ$ ($\phi < -150^\circ$).
One gets a false conclusion that the galaxy center
contains a decoupled disk whose inclination to the
sky plane is $15^\circ - 25^\circ$ smaller than that of the "outer"
disk of the galaxy. The real angle between the disk
planes can be as high as 90\degr~(the polar disk).

The point is that, although at $i > 50^\circ-60^\circ$, the $V_{obs}$
dependence on $PA$ in the equation (2) differs strongly from
the cosine relation, radial motions whose projected
velocities are maximum at the minor axis of the galaxy
distort the pattern implied by equation (2) making it
more like a cosinusoidal corresponding to smaller $i$. It is
thus evident that restricting the analysis to surface photometry
alone makes it rather difficult to distinguish a
minibar from a nuclear polar disk; one may erroneously
take the observed pattern for a manifestation of gas
moving in curricular orbits with smaller inclination
angles, whereas it might simply be a line-of-sight projection
of a superposition of circular and noncircular
motions in the bar. A turn of the dynamical axis should
also be observed in the case of a real tilted (including
polar) disk, and analyzing the velocity field alone can
lead to wrong conclusions. Recall that in the tilted disk
case, both the dynamical and photometric axes should
turn in the same direction (Zasov \& Moiseev 1999).
A tilted disk can therefore be distinguished from a bar
only by comparing the PA inferred from the velocity
field analysis with the surface photometry data.

\section{Kinematics of gas in the galaxy NGC~972}

In order to illustrate our approach, we applied it to
the NGC 972 galaxy. The  galaxy are observed in the
[NII]$\lambda  6583$ emission line with the scanning Fabry-Perot
Interferometer attached to the 6-m BTA telescope of the Special
Astrophysical Observatory of the Russian Academy
of Sciences. The seeing and spectral resolution
were 1\farcs5 and $\sim50\km$, respectively.
Two velocity fields with different image orientations
relative to the detector (two-dimensional
photon counter IPCS) are constructed. The results of the velocity-field
analysis
are briefly described by Zasov \& Moiseev (1999),
and a more detailed paper is now in preparation.
Fig. 4 shows the radial profiles of the position
angle of the dynamical axis $PA$ and inclination angle
$i$ formally inferred from the velocity field in terms of
the circular rotation approximation as described in Section
2. Dots with error bars are mean values averaged over both velocity fields.

The mean inclination of the galaxy to the line of
sight is $i\approx64^\circ$; the position angle of the line-of-nodes,
$PA = 150^\circ$, in agreement with the orientation of the outermost
K-band isophotes ($2.2 \,\mu$) based on observations
taken at the UKIRT infrared telescope (Hawaii).
The abrupt change of $PA$ and $i$ at $r < 10''$ can be
interpreted in terms of both inclined disk or minibar
models.

However, comparing the turns of the dynamic axis
with the K-band isophote orientation allows this ambiguity
to be resolved. The black line segment in Fig. 4
indicates the mean orientation in the central isophotes
of the galaxy, i.e., the position of the photometric axis.
It is evident from the figure that the orientation of the
photometric axis (relative to the line-of-nodes) varies in
"antiphase" with that of the dynamical axis, a pattern
that is indicative of a $\sim10''$ (1 kpc) bar whose axis
makes an angle of $\phi\approx-120^\circ$ to the line-of-nodes. Variations
of angle i in this case are due to noncircular
motions in the bar, as noted above in Section 3.

\begin{figure}
\psfig{figure=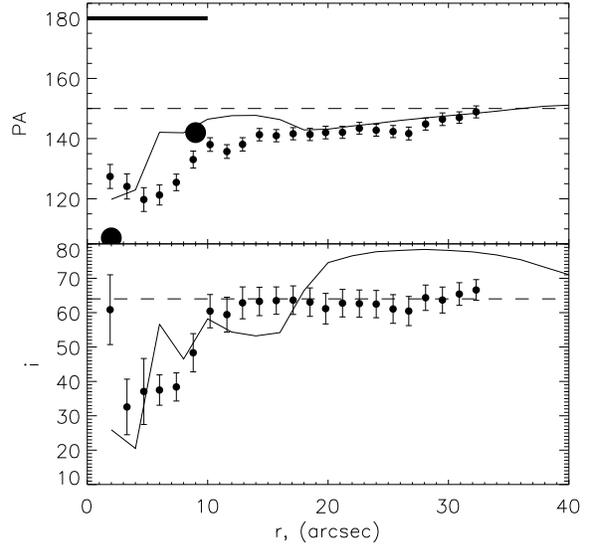,width=8 cm }%<---Nii
\caption{
Results of an analysis of the [NII] line velocity fields
of the NGC 972 galaxy (black dots with error bars). Two
large circles show the results of Zasov \& Sil'chenko
(1996). Above: radial variations of the dynamical axis orientation,
thick interval shows the mean orientation of central
K-band isophotes of NGC 972 (bar position). Below: gaseous
disk tilts as inferred from velocity fields. Solid lines in
each graph show model dependencies for the dimensionless
time instant $T = 2.2$ and bar turn angle $\phi = 120^\circ$.
}
\end{figure}

We constructed model velocity fields with orientation
parameters equal to those of the NGC 972 disk for
various time instants $T$ corresponding to a gradual
increase of the contrast of the gravitational potential of
the bar (Levy et al. 1996). Solid lines in Fig. 4 show the
results for the model velocity field at $T = 2.2$. Note that
the observed plots agree well with the computed ones
given the fact that, strictly speaking, the mass distribution
in NGC 972 differs from that of the adopted model.

\section{CONCLUSION}
Our analysis of model velocity fields constructed
from numerical nonlinear two-dimensional simulations
of gas flows in bars showed that, if the galaxy plane is
inclined by more than $i > 60^\circ-70^\circ$ to the sky plane, the
observed distribution of gas line-of-sight velocities in
the bar region can create an illusion of a "tilted nuclear
disk", implying a wrong interpretation of the observed
velocity field.

However, in the case of a bar, the dynamical axis
always turns with respect to the line-of-nodes in
antiphase with the photometric axis, thereby allowing
the bar to be distinguished from a tilted disk by a combined
analysis of the kinematics and photometry.
We illustrated the use of this criterion by applying it
to the NGC 972 galaxy.

\acknowledgements{
We are grateful to A.V. Zasov for calling our attention
to the problem and providing velocity fields of
NGC 972 obtained in the framework of his BTA observing
program. We are also grateful to O.K. Sil'chenko for
her critical comments, interest in our work, and primary
reduction of observational material. The image of the
NGC 972 galaxy was provided by S. Rider. We thank
the staff members of the SFVO laboratory (Special
Astronomical Observatory, Russian Academy of Sciences)
who performed all observations at the BTA telescope.
Special thanks are due to V.L. Afanas'ev for his
constructive criticism and valuable suggestions. The
work was partly supported by the Russian Foundation
for Basic Research (grant no. 98-02-17102) and INTAS
(grant no. 95-0988).}

\end{document}